\documentclass[aps,pre,showpacs,superscriptaddress,floatfix]{revtex4}
\usepackage{graphicx}
\def\w*{\omega_{*}}
\def\up{\omega_{*+}}
\def\dn{\omega_{*-}}
\def\beq{\begin{equation}}
\def\eeq{\end{equation}}
\def\bea{\begin{eqnarray}}
\def\eea{\end{eqnarray}}
\def\rt{\rightarrow}

\begin{document}

\title{Critical behavior of loops and biconnected clusters on fractals 
of dimension $d < 2$.}

\author{Dibyendu Das}
\affiliation{
Department of Physics, Indian Institute of Technology Bombay, 
Powai, Mumbai 400076, India}
\author{Supravat Dey}
\affiliation{
Department of Physics, Indian Institute of Technology Bombay, 
Powai, Mumbai 400076, India}
\author{Jesper Lykke Jacobsen}
\affiliation{Laboratoire de Physique Th\'eorique, Ecole Normale Sup\'erieure,
24 rue Lhomond, 75321 Paris cedex 05, France}
\affiliation{Institut de Physique Th$\acute{e}$orique, CEA Saclay, 
91191 Gif sur Yvette, France}
\author{Deepak Dhar}
\affiliation{Department of Theoretical Physics, 
Tata Institute of Fundamental Research, 
Homi Bhabha Road, Mumbai - 400005, India}

\date{\today}

\begin{abstract}
  We solve the O($n$) model, defined in terms of self- and mutually
  avoiding loops coexisting with voids, on a $3$-simplex fractal
  lattice, using an exact real space renormalization group technique.
  As the density of voids is decreased, the model shows a critical
  point, and for even lower densities of voids, there is a dense phase
  showing power-law correlations, with critical exponents that depend
  on $n$, but are independent of density.  At $n=-2$ on the dilute
  branch, a trivalent vertex defect acts as a marginal
  perturbation. We define a model of biconnected clusters which allows
  for a finite density of such vertices. As $n$ is varied, we get a
  line of critical points of this generalized model, emanating from
  the point of marginality in the original loop model.  We also study
  another perturbation of adding local bending rigidity to the loop
  model, and find that it does not affect the universality class.
 \end{abstract}


\maketitle

\section{Introduction}
\label{intro}

The loop model is a very important model in statistical physics.  It
was defined originally in terms of the high temperature expansion of
the $n$-vector model \cite{Mukamel,Nienhuis82,Baxter86}. The cases of
$n=0,1,2$ correspond to well-studied cases of self-avoiding polymers
\cite{deGennes}, the critical Ising and the XY models \cite{Zinnbook},
respectively. The model has been studied quite extensively in $d=2$
dimensions, in several variants, including fully packed or dilute
versions, and with loops of more than one type. One can determine the
critical exponents of the model on the hexagonal and square lattices
using the Bethe Ansatz technique, the Coulomb gas method, and
numerical techniques involving exact diagonalization of transfer
matrices of systems on finite width cylinders
\cite{Nienhuis82,Baxter86,Saleur86,Batchelor88,DupSal87,Nienhuis92,Kondev98}.
The critical behavior of the model is also related to the dimer model,
edge coloring model, compact polymer model (for $n \leq 2$)
\cite{Batchelor94,Kondev95,Kondev98}, and the hard hexagon model (for
$n > 2$) \cite{guo}.  In the presence of a staggered field, the model
gets related to the critical Potts model \cite{das}.

The loop model to be studied here is defined by the partition function 
\beq 
Z_{\rm loop} = \sum_{\cal C} n^{\cal L} \omega^{\cal V},
\label{Zloop}
\eeq where the summation is over configurations ${\cal C}$ of
self-  and mutually-avoiding loops on a lattice. In the above, $n$
is the weight of a loop, $\omega$ is the weight of a vacancy (i.e., a
lattice vertex not visited by any loop), ${\cal L}$ denotes the number
of loops in a given configuration, and ${\cal V}$ the number of
vacancies.

In Eq.~(\ref{Zloop}), large values of $\omega$ correspond to a small
average density of loops, and thus to the high-temperature phase of
the $n$-vector model.  As $\omega$ is decreased, the average density
of loops and the mean loop size increases, and diverges as $\omega$
tends to an $n$-dependent critical value $\up(n)$. The critical
behavior of the loop model at $\up$ gives the critical behavior of the
$n$-vector model at its critical point in $d \geq 2$. For $\omega
< \up$ and $n\leq 2$, the model shows a critical phase, which is
called the {\it dense phase}. In $d=2$, the critical exponents of the
dense phase have been determined exactly
\cite{Nienhuis82,Batchelor88,DupSal87,Nienhuis92}. These vary with   
$n$, but are independent of the precise value of $\omega$.

The critical behavior of the loop model, and exponents of the dense
phase have been less studied for $d \neq 2$. One expects that the
dense phase of the loop model for $d>2$ would be related to the the
low-temperature phase of the $n$-vector model. The latter shows
power-law correlations in the entire low-temperature phase because of
the existence of gapless Goldstone modes. In $d=2$, Jacobsen {\it
et.~al.} \cite{read} have shown that allowing loops to intersect or
not leads to different critical behavior, and the dense phase of the
loop model is different from the Goldstone phase.

\begin{figure}
\includegraphics[width=8cm]{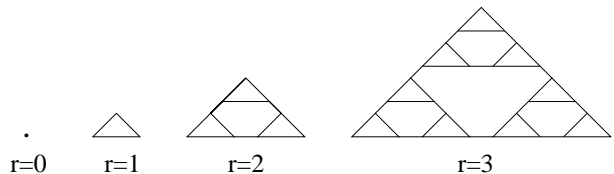}
\caption{The construction of the $3$-simplex lattice is shown from level
$r=0$ to $r=1$ and recursively for higher $r$.}
\label{3simplex}
\end{figure}

\begin{figure}
\includegraphics[width=12cm, height=3cm]{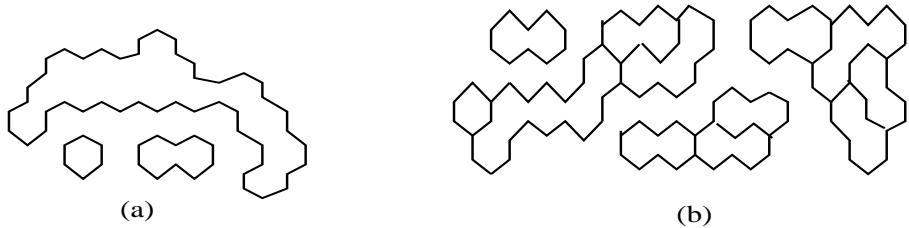}
\caption{(a) A typical configuration of the loop model.  (b) A typical
graph showing several biconnected clusters that contribute to the
partition function of the bicon clusters model.}
\label{bicon}
\end{figure}

It therefore seems interesting to study the loop model in dimensions
other than $d=2$.  In this paper, we study the loop model on fractal
lattices with finite ramification index. We will take the $3$-simplex
lattice (see Fig.~\ref{3simplex}) as the simplest example of this
type. The treatment is easily extended to other fractals. The loop
model shows a nontrivial critical behavior for $n \leq 1$ on this
fractal, and we determine the critical behavior near the dilute
critical point $\up$. We also study the critical properties of the
dense phase. Further we define a generalization of the loop model, to
be called the biconnected clusters model---or ``bicon clusters model'' 
for
short---in which we allow the summation in the partition function to
include all biconnected clusters, as shown in Fig.~\ref{bicon}.  A
cluster is called bi-connected, iff between any two points of the
cluster, there are at least two disjoint self-avoiding paths. Clearly
all simple loops are biconnected. It turns out that this changes the
critical behavior of the model, and we determine the new critical
exponents.

A $d=2$ variant of the bicon model is known as the ``net model''
\cite{Fendley}. This model is relevant for the study of quantum models
whose ground states are endowed with topological order.  More
generally, the constraint of $k$-connectedness has also been imposed
in classical models of clusters, such as percolation \cite{Zinn} and
the Potts model \cite{Deng}.

There have been several studies of critical behavior of statistical 
mechanical systems on finitely ramified fractals \cite{kenevic}. 
The Ising model was the earliest to be studied on fractals
\cite{nelson}.  It was followed by a study of self-avoiding polymers
on a $3$-simplex lattice \cite{dhar2}.  Later self-interacting
self-avoiding polymers \cite{klein}, and other trails \cite{Chang},
have been studied. The Lee Yang edge singularity for Ising model was
considered in \cite{kenevic1}. The collapse transition of branched
polymers was studied in \cite{kenevic2}. The distribution of sizes of
erased loops for loop-erased random walks was studied in
\cite{abhishek}.  In \cite{sumedha}, it was studied how the number of
self avoiding rings going through a site varies with the position of
the site. Polymers with bending energy were studied on $3$-simplex and
other fractal lattices in \cite{maritan}. For a recent
review, see \cite{dhar3}.  

The detailed plan of the paper is as follows. In Section
\ref{branches} we find the critical branches of the loop model, and in
section \ref{4expo} the exponents are calculated.  In section
\ref{net} we study the generalized ``bicon model'', as mentioned
above. 
In section \ref{bend} we study the
loop model with extra energies for local bending. In section
\ref{conclude} we summarize our results.

\section{The critical branches of the loop model}
\label{branches}

\begin{figure}
\includegraphics[width=12cm]{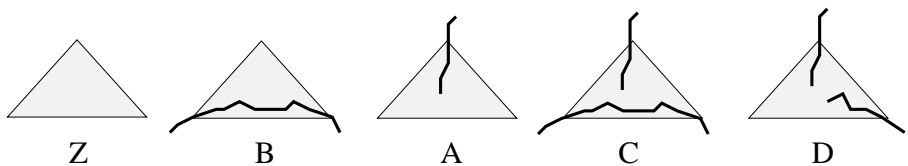}
\caption{The schematic representations of various restricted partition
functions $Z$, $B$, $A$, $C$ and $D$ for an $r$-th level triangle.}
\label{3simplex1}
\end{figure}

The recursive construction of the $3$-simplex fractal lattice,
through a series of levels $r$, is illustrated in Fig.~\ref{3simplex}.
Let $Z_r$ be the partition function at level $r$, i.e. the sum of 
statistical weights of all configurations with loops contained within 
level $r$, and denoted by an empty triangle in Fig.~\ref{3simplex1}. 
We may have loops which close at levels higher than $r$. 
Such a loop will come out of the corner vertices of an $r$-th 
level triangle. The restricted partition function for configurations with 
such an open chain passing out of the corner vertices of a $r$-th 
level triangle is denoted by $B_r$ and shown in Fig.~\ref{3simplex1}.

These functions $Z_r$ and $B_r$ can be recursively related to their 
counterparts at the $(r+1)$-th level by the following equations:   
\bea
Z_{r+1}={Z_r}^3 + n {B_r}^3 \\
\label{Z}
B_{r+1}=Z_r {B_r}^2 + {B_r}^3.  
\label{B} 
\eea
At the $0$-th level the lattice has just a single site. In that case 
$Z_0 = \omega$ is the weight of it being empty, and $B_0 = 1$ is the weight 
that a loop passes through it. If we define 
$\omega_r = Z_r/B_r$, then the recursion relation for the latter is 
\bea
\omega_{r+1}=\frac{\omega_{r}^{3} + n}{\omega_r + 1}.
\label{Y}
\eea
At the $0$-th level, $\omega_0 = Z_0/B_0 = \omega$. 
The total free energy on an infinite lattice is   
\bea 
F(n,\omega) &=& \lim_{r\rightarrow\infty} \frac{1}{3^r} ~ {\rm ln}~Z_{r+1}. 
\label{F1}
\eea
Eq.~(\ref{F1}) above may be rewritten, using 
$Z_{r+1} = Z_r^3 (Z_{r+1}/Z_r^3)$, and 
after a little algebra we obtain
\bea
F(n,\omega) = \frac{1}{3} {\rm ln}(1 + n {\omega}^{-3}) + \frac{1}{3} F\left(n,\frac{\omega^{3}+n}{\omega + 1}\right), 
\label{F2}
\eea which is the usual form \cite{Cardy} of the recursion relation
for the free energy under the real-space renormalization group
(RG). Apart from $\omega_* = \infty$, the finite-valued fixed points
of the RG flow (\ref{Y}) are the roots of the following equation:
\begin{equation}
\w*^3-\w*^2-\w*+n=0.
\label{fixed} 
\end{equation}
The special case  $n =  0$  corresponding to self-avoiding 
polymers was  studied earlier \cite{dhar2}. 

\begin{figure}
\includegraphics[width=8cm,angle=0]{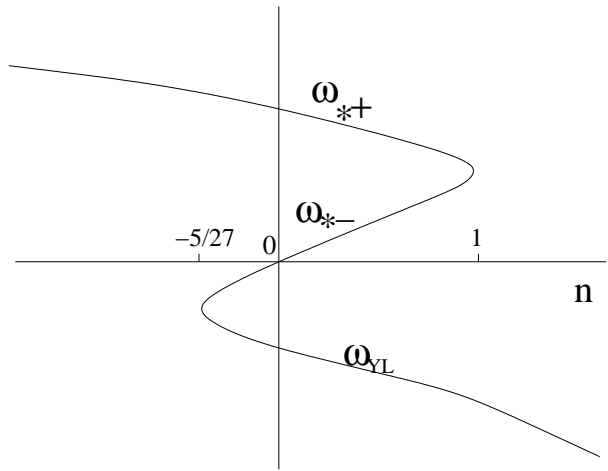}
\caption{A schematic picture of the three lines of fixed point 
$\up$, $\dn$ and $\omega_{YL}$ plotted against $n$. 
}
\label{branch}
\end{figure}

Eq.~(\ref{fixed}) has three real solutions for  
$-5/27 \leq n \leq 1$, and only one real solution for the regions $n>1$ and 
$n < -5/27$ (see Fig.~\ref{branch}). The three fixed points for the region  $-5/27 
\leq n \leq 1$
are given by the formulae  
\begin{eqnarray}
\omega_* &=& \frac{1}{3} \left( 1 + 4 \cos \left( \frac{\theta}{3}
\right) \right), \label{w+}\\
n &=& \frac{1}{27} \left( 11 - 16 \cos(\theta) \right), \label{theta}
\end{eqnarray}
where $\theta \in [0,3\pi]$. The three solutions correspond to the
following subdivision of the parameter range of $\theta$:
\begin{itemize}
\item For $\theta \in [0,\pi]$, $n$ runs from $-5/27$ to $1$, and
$\omega$ from $5/3$ to $1$. We shall refer to this line of fixed points 
as the {\it dilute branch}
and denote it by $\up(n)$. It is physical ($n \ge 0$) for 
$\theta \ge \cos^{-1}(11/16) \simeq 0.8128$. 
\item For $\theta \in [\pi,2\pi]$, $n$ runs from $1$ to $-5/27$, and
$\omega$ from $1$ to $-1/3$. We shall refer to this as the {\it dense branch}
and denote it by $\dn(n)$. It is physical ($n \ge 0$ and $\omega \ge 0$) for
$\theta \le 2\pi-\cos^{-1}(11/16) \simeq 5.4704$.
\item For $\theta \in [2\pi,3\pi]$, $n$ runs from $-5/27$ to $1$, and
$\omega$ from $-1/3$ to $-1$. 

This unphysical critical point at negative $\omega$ may be called 
the Yang Lee edge singularity in our problem. We shall denote it by 
$\omega_{YL}$.  However, we note that the critical behavior at this 
"Yang Lee edge singularity"  is not independent of $n$.

\end{itemize}

The three branches $\up$, $\dn$ and $\omega_{YL}$ are shown in
figure \ref{branch}. Note that unlike the two-dimensional loop model,
where the critical region between the dilute and dense branches
extends up to $n=2$, the domain of criticality of the present model is
smaller. But here too, the upper branch is a repulsive fixed
line. Starting from any $\omega > \up$ the RG flows take one to
$\omega = \infty$ while starting from any $\omega < \up$ the lower
branch $\dn(n)$ serves as a line of attraction of the RG flow. We will
refer to the region with $\omega < \up$ as the {\it dense phase}.

For $n<-5/27$, the dilute branch can be extended backwards by
replacing $\theta = i {\tilde \theta}$ (with ${\tilde \theta} \in
[0,\infty)$), i.e replacing ``$\cos$'' by ``$\cosh$'' and $\theta$ by
${\tilde \theta}$ in Eqs.~(\ref{w+}) and (\ref{theta}).  Although this
regime is unphysical, we will show in section \ref{3string} that the
point $(n=-2,\omega=2)$ on it is of some signficance.

For $n > 1$, the nontrivial fixed point other than $\w* = \infty$ is
negative, and is obtained by replacing $\theta = 3\pi + i {\tilde
\theta}$ (with ${\tilde \theta} \in [0,\infty)$), i.e. replacing
``$\cos$'' by ``$-\cosh$'' and $\theta$ by ${\tilde \theta}$ in
Eqs.~(\ref{w+}) and (\ref{theta}). There is no nontrivial critical
point for positive $\omega$. It can be shown
\cite{supravat} that the loop model for $n = 2$ can be exactly mapped
to a weighted $6$-vertex model, which also can have  no nontrivial 
critical point for positive weights in this regime.

\section{The exponents of the loop model}
\label{4expo}

In two dimensions, for a generic point lying on the dilute branch or
in the dense phase, the loop model is known to have an infinite set of
exponents characterizing its critical behaviour
\cite{Kondev95,saleur}.  The full set of principal exponents consists
of an infinite number of correlation exponents associated with defects
called the $k$-string defects, plus one thermal exponent.

A $k$-string defect has an integer $k (\geq 1)$ number of ``open
chains'' originating from a local neighborhood and terminating in an
anti-defect very far away. Since chains are disallowed in the original
loop model, such structures may be imagined to be created in isolation,
as defects, upon application of suitable small external
fields. A closed loop with {\it two points marked} on
it is equivalent to a $2$-string defect, and this observation can be 
used
\cite{Kondev95,saleur,das} to find the fractal dimension of the closed
loops in the loop model. Although there is no upper bound on $k$ for
ordinary Euclidean lattices, on fractals due to finite ramification,
the exponent spectrum becomes finite. For the present model on the
$3$-simplex fractal lattice we can have such defects only with $k \leq
3$. [On the Sierpinski gasket, one could have $k \leq 4$.] 

Apart from the $k$-string defects, the loop model also has the thermal
defect and different exponents associated with it, namely the specific heat 
exponent, the correlation length exponent, etc. Among the latter only 
one is independent and others can be related to it by identities. Thus 
there are four independent exponents for the loop model on the 
$3$-simplex lattice, three $k$-string exponents plus one thermal exponent, 
and we calculate them in the following subsections \ref{thermal}-\ref{3string}.

\subsection{The exponents related to thermal excitation}
\label{thermal}

Let us define $(\omega - \up)$ as the ``thermal distance'' from 
the critical point $\up$. Starting from $\omega_0 = \omega > \up$, under 
succesive iteration of Eq.~(\ref{Y}), the separation $\delta_r = (\omega_r 
- \up)$ increases with increasing $r$. We note that 
$\delta_0 = (\omega - \up)$. 
Using Eqs.~(\ref{Y}) and (\ref{fixed}), we easily find that 
$\delta_r = \lambda_T^{r} \delta_0$ with 
\bea
\lambda_T = \left(\frac{3 - \up^{-1}}{1 + n \up^{-3}} \right).
\label{lambdaT}
\eea
The thermal exponent $y_T$ is defined as usual by the fact that 
at the $r$-th step of RG transformation, the thermal distance
$\delta_r \sim (2^{y_T})^{r} \delta_0$. Thus comparing the two expressions for 
$\delta_r$ we obtain 
\bea
y_T = \frac{\ln{\lambda_T}}{\ln{2}}
\label{yT}
\eea 

The specific heat exponent $\alpha$ 
may be defined via the scaling behavior of the non-analytic part of the 
free energy, namely $F \sim (\omega - \up)^{2-\alpha}$. Substituting
the latter scaling form into Eq.~(\ref{F2}) and comparing the singular 
terms, a straightforward  calculation yields
\bea
\alpha = 2 - \frac{\ln 3}{\ln \lambda_T},
\label{alpha+}
\eea
where $\lambda_T$ is given by Eq.~(\ref{lambdaT}). 

Thermal fluctuations create thermal defects and the correlation function 
$C({\bf R})$ for two such defects separated by distance $|{\bf R}|$ 
is defined for $\omega > \up$ as 
\bea
C({\bf R}) = \frac{1}{|{\bf R}|^{c}} f\left(\frac{|{\bf R}|}{\xi}\right).
\label{corrT}
\eea
The correlation length $\xi$ is finite for $\omega > \up$ and the 
scaling relation $\xi \sim (\omega - \up)^{-\nu}$ defines the 
correlation length exponent $\nu$. Under RG, if $\xi \sim b$, since 
$(\omega - \up) \sim b^{-y_T}$ it follows immediately that 
\bea
\nu = 1/y_T = \frac{\ln 2}{\ln \lambda_T}.
\label{nu}
\eea 
For $\omega = \up$, in Eq.~(\ref{corrT}), $\xi \rt \infty$ and $c
= 2 x_T$, where $x_T$ is called the thermal correlation exponent at
criticality.  From standard RG arguments \cite{Cardy} it can be shown
that $x_T = d - y_T$, where 
\bea
d = \frac{\ln 3}{\ln 2}
\label{d} 
\eea
is the box dimension
of the $3$-simplex fractal lattice.  From Eqs.~(\ref{alpha+}) and
(\ref{nu}) we see that the hyperscaling relation $d \nu = 2 - \alpha$
holds.

Apart from the exponents of thermal origin discussed above, 
there are several other  exponents caracterizing the behavior of other 
physically interesting observables, which  can be expressed in terms of 
the exponents described above: 
\begin{itemize}
\item The fractal dimension of the loops given by $d_{f}$ relates the
length $s$ and radius $R$ of a loop as $s \sim R^{d_f}$.  The exponent
$d_f$ is distinct for the two critical regions $\omega = \up$ (dilute
branch) and $\omega < \up$ (dense phase). We show in section
\ref{2string} that they can be easily related to the $2$-string
correlation exponent $x_2$. 
\item For $\omega > \up$, the distribution of
loop sizes $s$ has an exponential cutoff $\sim \exp( - a s)$.  The
cutoff length $a^{-1} \sim \xi^{d_f}$, where $\xi$ is the finite
correlation length in the disordered phase. One can find $a^{-1}$ 
as in the derivation of $\nu$ above. 
\item For the critical regimes 
$\omega = \up$ and $\omega < \up$, the probability distribution
of loop size $s$ is an unbounded power law $\sim s^{-({\tau - 1})}$.
The exponent $\tau$ is distinct for $\omega = \up$ and $\omega < \up$,
and it is easy to express the exponent $\tau$ in terms of the exponents
$x_2$ and the fractal dimension $d_f$ of the loops (see section
\ref{2string}).  
\end{itemize}
The exponents like $d_f$ and $\tau$ can all be expressed 
in terms of the ``$k$-string defect'' exponents which we 
derive in the following three subsections.

\subsection{A $1$-string defect}
\label{1string}

Let the application of a small external magnetic field $h_1$ create
an open chain, or a $1$-string defect, with the magnetic scaling exponent 
$y_1$ associated with it. Assuming for $h_1 = 0$ the system is critical,
a finite $h_1$ introduces a finite correlation length $\xi$ given by 
\bea
\xi(h_1) \sim h_1^{-{1/y_1}}.
\label{y1} 
\eea
Eq.~(\ref{y1}) defines the scaling exponent $y_1$.  
In real space RG, upon coarse-graining from level $r$ to level $(r+1)$,
$\xi$ increases by a factor of $2$, while $h_1$ increases by a factor 
of $\lambda_1$, i.e.  $\xi(h_1) = 2 \xi(\lambda_1 h_1)$. Below, we 
will calculate $\lambda_1$, and using this with Eq.~(\ref{y1}), one gets 
\bea 
 y_1 = \frac{\ln \lambda_1}{\ln 2}.
\label{y11} 
\eea 

The probability $G({\bf X - Y})$ \cite{saleur} of $k$ strings originating 
at ${\bf X}$ and ending at ${\bf Y}$ in the critical phase is given by 
\bea
G({\bf X - Y}) \sim 1/|{\bf X} - {\bf Y}|^{2 x_k}.
\label{Gk}  
\eea 
>From standard RG analysis it is known \cite{Cardy} that the correlation 
exponent $x_k$ is related to $y_k$ via the relation:
\bea
x_k = d - y_k.
\label{xk} 
\eea   
Thus for our $1$-string defect, by using Eq.~(\ref{y11}) and box dimension
(\ref{d}), we get
\bea
x_1 = d - y_1 = \frac{\ln(3/\lambda_1)}{\ln 2}.
\label{x1}
\eea

Now we proceed to obtain the scale factor $\lambda_1$. First we note
that an $1$-string defect may have its one endpoint inside an $r$-th
level triangle. The corresponding restricted partition functions $A_r$
and $C_r$ are represented by diagrams $A$ and $C$ in
Fig.~\ref{3simplex1}.  The recursion relations for ${\tilde A}_{r} =
A_r/Z_r$, ${\tilde C}_{r} = C_r/Z_r$, and ${\tilde D}_{r} = D_r/Z_r$
are as follows:
\bea {\tilde A}_{r+1} &=& \frac{{{\tilde
      A}_r}(1+2\omega^{-1}_r+2{\omega^{-2}_r})+{{\tilde
      C}_r}{\omega^{-2}_r}(n+2)}{1+n {\omega^{-3}_r}}
\label{A} \\
{\tilde C}_{r+1} &=& \frac{({\tilde A}_r+3{\tilde C}_{r}){\omega^{-2}_r}}{1+n{\omega^{-3}_r}} 
\label{C} \\
{\tilde D}_{r+1} &=&\frac{{{\tilde A}_r}^2+{{\tilde C}_r}^2 \omega^{-1}_r(n+6)+ 4 {\tilde A}_r \omega^{-1}_r {\tilde C}_r+ 2 {{\tilde A}_r}^2 \omega^{-1}_r+ {\tilde D}_r (2\omega^{-1}_r+3{\omega^{-2}_r}) }{1+n{\omega^{-3}_r}}
\label{D}
\eea

We can linearize Eqs.~(\ref{A})--(\ref{C}) around the fixed 
point $(A^*,C^*,D^*) =(0,0,0)$, and we are left with the matrix 
\bea  
\left[ \begin{array}{cc}
 a+2        & ~~a(n+2)\w*^{-2}  \\
~ & ~ \\
 a\w*^{-2}  & ~~3a\w*^{-2}       \\ 
\end{array} \right]
\eea 
with $a = 1/(1+n\w*^{-3})$. The largest eigenvalue of the matrix 
is $\lambda_1$ and is given by 
\bea 
\lambda_1 = \frac{1}{2\w*^2}\left(3a +\w*^2(a+2) +
\sqrt{a^2(17+4 n) - 6 a\w*^2(a+2)+\w*^4(a+2)^2}\right).
\label{lambda1}  
\eea 
Note that in Eq.~(\ref{lambda1}), $\w*$ must be replaced by
$\up$ or $\dn$ in order to describe the dilute or dense branch, respectively.
Using Eq.~(\ref{lambda1}), we may thus write 
${\tilde A}_r \sim \lambda_1^{r} {\tilde A}_0$. 
It is clear that $A_0$ is proportional to the small 
external field $h_1$, and thus using Eq.~(\ref{lambda1}),
we may read off the values of $y_1$ and $x_1$ in
Eqs.~(\ref{y11}) and (\ref{x1}).

Another exponent of interest is $\gamma_1$ associated with 
the approach from above of the dilute critical branch. It is 
defined via the scaling behavior of the average open chain length 
$\langle l_1 \rangle \sim (\omega - \up)^{-\gamma_1}$. 
In general,  
\bea
\langle l_1 \rangle = \lim_{N \rightarrow \infty} \frac{1}{N} \sum_{n=1}^{\infty} {c_n p_n}, 
\eea
where $c_n$ is the number of distinct configurations each with an open chain 
length $n$ in a lattice of size $N$, and $p_n$ is the relative weight factor 
for such a configuration normalized by the partition function. On our 
fractal lattice, this becomes   
\bea
\langle l_1 \rangle & = & \sum_{r=1}^{\infty}
\frac{1}{3^r}\left(\frac{f_r}{Z_r}\right) \nonumber \\
& = & \sum_{r=1}^{\infty} \frac{1}{3^r} \frac{\left(3 {\tilde A}_{r-1}^2 + 3 \omega^{-1}_{r-1} {\tilde A}_{r-1}^2 + 3 \omega^{-2}_{r-1} {\tilde D}_{r-1} \right)}{\left( 1 + n \omega^{-3}_{r-1} \right)},  
\label{f_1}
\eea  
where the statistical weight of fully containing an open chain at the 
$(r+1)$th level is $f_{r+1} = 3 A_r^2 Z_r + 3 B_r A_r^2 + 3 B_r^2 D_r$. 
For a fixed and small  distance from the critical point 
$\delta_0 = (\omega - \up)$,
since $\langle l_1 \rangle$ is finite, 
the sum in Eq.~(\ref{f_1}) is sharply cut off at some finite level $r = r_0$.
We note that for $r > r_0$,  ${\tilde A}_{1,r} \sim \lambda_1^{r_0}$, and 
${\tilde D}_{r} \sim \lambda^{2r_0}$, and $\delta_0 \sim 
({\rm constant})/\lambda_T^{r_0}$ (see Eq.~(\ref{lambdaT})). 
It immediately follows from the scaling relation 
$\langle l_1 \rangle \sim {\delta_0}^{-\gamma_1}$ that 
\begin{equation}
\gamma_1= \frac{\ln({\lambda^2}/3)}{\ln \lambda_T}.
\label{gamma+}
\end{equation}
As a check of consistency, we may verify the following exponent
equality \cite{saleur}, using Eqs.~(\ref{gamma+}), (\ref{x1}) and (\ref{nu}),
\bea 
\gamma_1 = (d - 2 x_1) \nu.
\label{gamma1}
\eea

\subsection{A $2$-string defect}
\label{2string}

Let a triangle at the $r$-th level which has endpoints of $2$-strings 
on two neighboring sites inside it, be defined to have a statistical 
weight $D^{'}_r$. Note that this weight is
different from $D_r$ (see section \ref{2string} above), since 
the latter puts no restriction on the location of the endpoints. 
Further defining ${\tilde D}^{'} = D^{'}/Z$, we easily see that its 
recursion equation is 
\bea 
{\tilde D}^{'}_{r+1} = \frac{2 \omega^{-1}_r + 3
\omega^{-2}_r}{1 + n \omega^{-3}_r}~{\tilde D}^{'}_r.
\label{D'}
\eea 
Replacing $\omega_r$ with the fixed point value $\w*$ in the above equation, 
we get ${\tilde D}^{'}_r \sim \lambda_2^{r} {\tilde D}^{'}_0$, with
\bea
\lambda_2 = \frac{2 {\w*}^{-1} + 3 {\w*}^{-2}}{1+n{\w*}^{-3}}  
          = \frac{2 {\w*} + 3}{\w* + 1},
\label{lambda2}
\eea  
where we have used Eq.~(\ref{fixed}).
Note again that in Eq.~(\ref{lambda2}) for $\lambda_2$ we have to use
$\w* = \up$ for the upper branch, and $\w* = \dn$ for the 
lower branch. 
If a small field $h_2$ creates a $2$-string defect 
and the finite correlation length arising due to it is 
$\xi(h_2) \sim h_2^{-{1/y_2}}$, we find that the scaling exponent
$y_2$ for a $2$-string defect is 
\bea 
y_2 = {{\rm ln}\lambda_2}/{{\rm ln}2}. 
\label{y2}
\eea
>From Eq.~(\ref{xk}) we conclude that the corresponding correlation 
exponent $x_2$ is given by 
\bea
x_2 = d - y_2.
\label{x2}
\eea

It is important to note that $y_2$ is also the {\it fractal dimension} 
$d_f$ of the
loops. This is explicitly seen as follows. 
Let $l_r$ be the typical length of a segment of a loop that goes from one 
corner vertex to another in one $B$-type triangle 
(see Fig. \ref{3simplex1}) of order $r$. Then the typical length 
of a loop which closes at the $(r+1)$-th level is $s_{r} = 3 l_r$.  
The recursion for $l_r$ is:
\bea 
l_{r+1} = \left(\frac{2 Z_r
B_r^2 + 3 B_r^3}{Z_r B_r^2 + B_r^3}\right) l_r.
\label{lr}  
\eea 
The above Eq.~(\ref{lr}) follows from the fact that at level
$(r+1)$, a typical loop length $l_{r+1}$ 
can be made of two $l_r$ segments with statistical weight $Z_r
B_r^2/B_{r+1}$, or three $l_r$ segments with statistical 
weight $B_r^3/B_{r+1}$ \cite{vani_df}. On the other hand, the typical 
diameter $R_r$ of such a segment of length $l_r$ 
is given by $R_r \sim 2^r$. 
Combining these two results and using the definition of $d_f$ given by the 
scaling behavior $s_r \sim l_r \sim R_r^{d_f}$, we conclude that   
\bea 
d_f = \frac{\ln \lambda_2}{\ln 2} = y_2.
\label{df} 
\eea 

We note that $d_f$, through $\lambda_2$, is distinct for $\w* = \up$
(dilute loops) and $\w* = \dn$ (dense loops).  For $n =0$ and in 
the dense loop phase, 
$\lambda_2 = 3$, and so $d_f$ coincides with the box dimension $d$ of
Eq.~(\ref{d}). This was to be expected, since the limit of dense loops
($\dn \to 0$) in fact means that there is a single loop covering the
entire lattice, i.e., the loop is Hamiltonian. For dilute loops, we
have $\up = \frac12(1+\sqrt{5})$, whence $\lambda_2 = \frac12(7-\sqrt{5})$
and $d_f \simeq 1.2522$.

The probability distribution $P(s)$ of 
loop size $s$ is an unbounded power law $\sim s^{-({\tau - 1})}$ for 
both the dilute branch and the dense phase. The exponent $\tau$
can be related to $x_2$ and $d_f$ following a general derivation 
as in \cite{kondev_loop}. Let $G_s({\bf R})$ be the probability 
that two points separated by ${\bf R}$ are on a loop of size $s$. 
The expected scaling form of $G_s({\bf R})$ is:
\bea
G_s({\bf R}) \sim s^m |{\bf R}|^{-c_1} f_1\left(\frac{|{\bf R}|}{s^{1/d_f}}\right).
\eea  
Firstly, the sum of $G_s({\bf R})$ over all the space, i.e. $\int d^d{\bf R} 
G_s({\bf R})$, is nothing but number of points on the loop $= s$. 
This gives a relation $d - c_1 = d_f (1 - m)$. 
Secondly, by definition (see Eq.~(\ref{Gk})) we have $G({\bf R}) = \int ds P(s) 
G_s({\bf R}) \sim |{\bf R}|^{-2 x_2}$. The latter relation combined with the 
former gives 
\bea
d_f (3 - \tau) = d - 2 x_2, 
\label{tau_s}
\eea
which is the desired result. 

\subsection{A $3$-string defect:}
\label{3string}

Let a triangle at the $r$-th level which has endpoints of $3$-strings 
on three neighbouring sites inside it, be defined to have a statistical 
weight $E^{'}_r$. The recursion relation of ${\tilde E}^{'} = E^{'}/Z$ is
\bea 
{\tilde E}^{'}_{r+1} = \frac{3 \omega^{-2}_r}{1 + n \omega^{-3}_r}~{\tilde
  E}^{'}_r. 
\label{E'}
\eea 
Assuming that a small field $h_3$ creates a $3$-string defect and 
the corresponding finite correlation length $\xi(h_3) \sim
h_3^{-{1/y_3}}$, we find that
\bea 
y_3 = {{\rm ln}\lambda_3}/{{\rm ln}2},
\label{y_3}  
\eea
where 
\bea
\lambda_3 = \frac{3 \w*}{\w*^3 + n}.
\label{lambda3}
\eea 
The correlation exponent $x_3$ corresponding to $y_3$ 
is given by Eq.~(\ref{xk}) with $k = 3$. 

We now note something very interesting. If we extend the upper dilute
critical branch to negative values of $n < -5/27$, 
we find that at the special point
$(\w* = 2, n = -2)$, we have $\lambda_3 = 1$ and $y_3 = 0$ (see
eqs. (\ref{y_3}) and (\ref{lambda3})). Thus although the $1$-, $2$-
and $3$-string defects are relevant for general $n$ on both the
critical branches, at the point $(\w* = 2, n = -2)$ the $3$-string
defect becomes marginal. In the next section \ref{net}, we define a
more general model called the ``biconnected cluster model'' 
which allows for
vertices of degree 3, each occuring with a finite weight $\omega_3$.
We find that the latter model has a new critical
line in its larger parameter space, which precisely meets   with the 
line of critical points corresponding to $\omega_3 = 0$ at the
point $(\w* = 2, n = -2)$ on the extended  dilute branch.

\section{The biconnected cluster model}
\label{net}

\begin{figure}
\includegraphics[width=9cm]{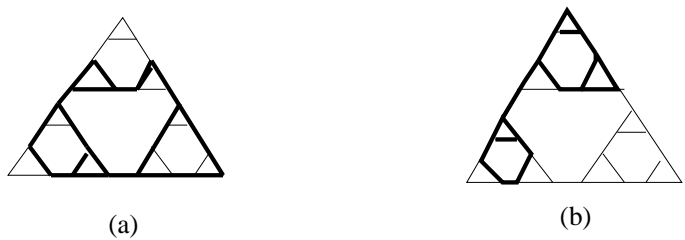}
\caption{(a) The structure with thick dark lines is a possible cluster
in the bicon clusters  model. (b) This is a disallowed cluster as it can 
be
disconnected into two pieces by cutting one link.}
\label{net0}
\end{figure}

The finding in section \ref{3string} that $y_3 = 0$ at a point in the
$(\omega,n)$ parameter space, motivates our study of a model called
the {\it biconnected cluster model} in which the allowed vertex
degrees are 0, 2 and 3. In other words, $3$-strings are allowed to
emerge from any vertex (see Fig.~\ref{net0}(a)). The connected
components are further required to be 2-connected, i.e., they cannot
be disconnected upon cutting a single link (see
Fig.~\ref{net0}(b))---note that this is a stronger requirement than
simply disallowing vertices of degree 1. Henceforth we will refer to
this model as the ``bicon clusters model'',  to distinguish it from the 
``loop model''
studied this far in the paper. A similar model---with no requirement
of biconnectedness---was studied in two dimensions in \cite{Fendley}
under the name of ``net model''. The partition sum for the bicon model
is \beq Z_{\rm bicon} = \sum_{\cal C} n^{\cal L} \omega^{\cal V}
\omega_3^{\cal U},
\label{Znet}
\eeq where the summation is over configurations ${\cal C}$ of
any number of self-, and  mutually-avoiding biconnected clusters.  Here 
$n$ is
the weight of a cluster of any size, $\omega$ is the weight of an
empty vertex, and $\omega_3$ is the weight of a vertex of degree 3.
Further, ${\cal L}$ denotes the number of clusters, ${\cal V}$ the
number of vacancies, and $\cal U$ the number of vertices of degree
$3$, in a given configuration.  Note that by setting $\omega_3 = 0$,
the bicon clustes model reduces to the usual loop model. Some of the
configurations that we exclude from the model (see Fig.~\ref{net0}(b)
for an example) will be treated as defect configurations called
``$k$-defects'' in section \ref{netstring} below.

\begin{figure}
\includegraphics[width=9cm]{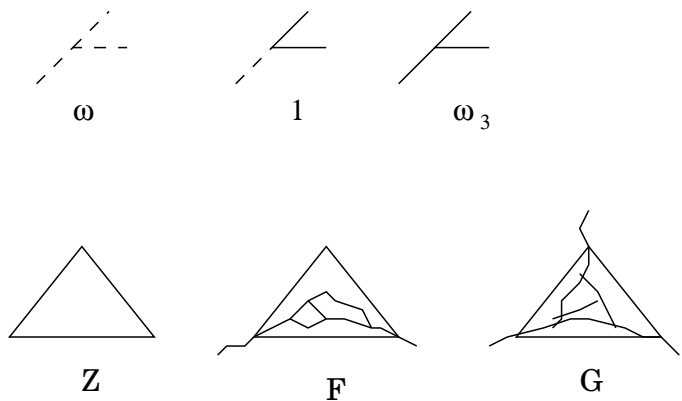}
\caption{The three vertex weights for the bicon clusters  model are shown 
in the
top row.  In the bottom row, the three schematic representations of the
restricted partition functions are shown.}
\label{net3}
\end{figure}   
The three possible vertex configurations with weights $\omega$, $1$
and $\omega_3$ are shown in Fig.~\ref{net3}.  Note that if ${\cal
  N}_k$ is the number of vertices of degree $k$, we have the simple
topological identity $2 {\cal N}_2 + 3 {\cal N}_3 = 2 L$, where $L$ is the total
number of links in the configuration. Therefore ${\cal N}_3 = {\cal U}$ is
necessarily even. Accordingly, the bicon clusters model partition 
function is a
function of $(\omega_3)^2$.  At the $r$-th level, the real space RG
closes for three restricted partition functions schematically shown as
$Z$, $F$ and $G$ in Fig.~\ref{net3}. While $Z_r$ is the
partition function summing over configurations with no strings coming
out of the corner vertices, $F_r$ and $G_r$ have two and three strings
coming out of the corners, respectively. Note that the constraint of
biconnectedness implies that it is not possible to have a
configuration with one string coming out of the corner vertices. At
the level $r = 0$, \bea Z_0 = \omega; ~~ F_0 = 1; ~~ G_0 = \omega_3.
\eea

At the $r$-th level, the recursion relations are: 
\bea 
Z_{r+1} &=& Z_r^3 + n F_r^3 \\ 
F_{r+1} &=& F_r^3 + F_r^2 Z_r + F_r G_r^2 \\ 
G_{r+1} &=& G_r^3 + 3 G_r F_r^2 
\eea 
Further defining ${\tilde F}_r = F_r/Z_r$, and ${\tilde G}_r = G_r/Z_r$, 
we get \bea
{\tilde F}_{r+1} &=& \frac{{\tilde F}_r^3 + {\tilde F}_r^2 + {\tilde
F}_r {\tilde G}_r^2}{1 + n {\tilde F}_r^3}
\label{F}\\
{\tilde G}_{r+1} &=& \frac{ {\tilde G}_r^3 + 3 {\tilde G}_r {\tilde F}_r^2}
{1 + n {\tilde F}_r^3} 
\label{G}
\eea Note that ${\tilde F}_0 = \omega^{-1}$ and ${\tilde G}_0 =
\omega_3 \omega^{-1}$. There are several fixed points $({\tilde
  F}_*,{\tilde G}_*)$ of Eqs.~(\ref{F})--(\ref{G}). Apart from the
trivial weak coupling fixed point $(0,0)$ and the fixed point
$(\infty,\infty)$ corresponding to all bonds being fully covered,
there are the dilute and dense fixed points of the loop model which we
denote by $({\tilde F}_{+},0)$ and $({\tilde F}_{-},0)$,
respectively. But most importantly there is a nontrivial fixed point
$({\tilde F}_*,{\tilde G}_*)$ given by, \bea {\tilde F}_* = 1/2 ~~~
{\rm and} ~~~ {\tilde G}_* = \sqrt{\frac{1}{4} + \frac{n}{8}}.
\label{newline}  
\eea The latter defines a new critical line in the
$(\omega,n,\omega_3)$ space, which terminates on one end at the point
$(\omega = 2, n = -2, \omega_3 = 0)$; note that this is the point
where we found $y_3$ to be marginal for the loop model in section
\ref{3string}. The dense-phase fixed point $({\tilde F}_{-},0)$ is
unstable to introduction of trivalent sites, and the limiting
behaviour of the critical net model is governed by the fixed point
$({\tilde F}_*,{\tilde G}_*)$.

If the starting value of ${\tilde F}_r$ and ${\tilde G}_r$ is near 
${\tilde F}_*$ and ${\tilde G}_*$, but a bit larger, it is easy to check 
that ${\tilde F}_r$ and ${\tilde G}_r$ diverge to infinity as $r$ and 
$r^{3/2}$ respectively. This implies that in the dense phase of the 
bicon clusters models, corner sites of triangles of high order belong to 
the infinite cluster with  a large probability, and this probability 
tends to $1$ as $r$ tends to infinity.

\subsection{The fractal dimension  of the biconnnected 
clusters}
\label{netdf}

Just like in the loop model, we
would like to find the fractal dimension of the clusters in the bicon 
clusters model at its critical point.
Consider two {\it marked} points $A$ and $B$ on a cluster, free of tadpole
overhangs. Imagine that these marks are ``defects'' created by
some external field $h$. In addition to $F$ and $G$, we now define new
functions $F^m$ and $G^m$ which are analogous to $F$ and $G$, except
that they represent configurations in which there is exactly one
marked point. At level $r = 0$ we have:
\bea
F_0^m &=& h \nonumber \\
G_0^m &=& h \omega_3. 
\eea 
The recursion relations are as follows: 
\bea 
F^m_{r+1} &=& 2 F_r F^m_r Z_r + 3 F^2_r F^m_r + 2 F_r G_r G^m_r + G^2_r F^m_r \\ 
G^m_{r+1} &=& 6 F_r F^m_r G_r + 3 F^2_r G^m_r + 3 G^2_r G^m_r.  
\eea 
The largest
eigenvalue of the $2 \times 2$ matrix obtained by linearizing the
above equations around the fixed point of Eq.~(\ref{newline}) is 
\bea
\lambda_2^{\rm bicon} = \frac{2(7+n) + \sqrt{n^2 + 20n + 52}}{n+8}, 
\eea 
and the fractal dimension of clusters on the new critical line is  
\bea 
d_f^{\rm bicon} = {\rm ln}(\lambda_2^{\rm bicon})/{{\rm ln} 2}.  
\eea 
The scale factor $\lambda_2^{\rm bicon} \rt \lambda_2 = 
7/3$ and thus $d_f^{\rm bicon} \rt d_f$, the fractal dimension of the
loops (see Eq.~(\ref{df})), at the point $(\omega = 2, n = -2, \omega_3 =
0)$, as expected.

\subsection{The $k$-defects}
\label{netstring}

\begin{figure}
\includegraphics[width=14cm]{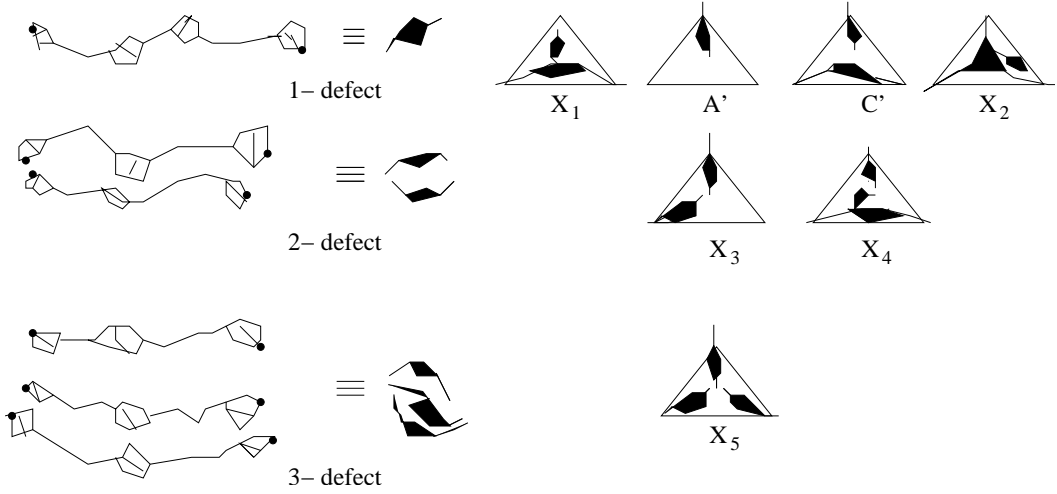}
\caption{The $1$-, $2$- and $3$-defects
are shown along with their shorthand representations (filled black
blobs).  The restricted partition functions contributing to the 
three defects are shown schematically on the right.}
\label{net4a}
\end{figure}

What kind of defects are natural extensions of $k$-string defects,
appropriate to the net model? A possibility is what we call the
``$k$-defects'' shown in Fig.~\ref{net4a}.  A $1$-defect has
multiple clusters connected in a series by strings, and the two dangling ends
are marked; they are depicted in short-hand as filled black blobs
(see Fig.~\ref{net4a}). The obvious motivation for defining such defects
is that as $\omega_3 \rightarrow 0$, they become our usual
``$k$-string'' defects in the loop model.  The restricted partition
functions contributing to each $k$-defect is shown
alongside the defects in Fig.~\ref{net4a}.

{\it $1$-defect dimension}: In Fig.~\ref{net4a}, $A^{'}$, $C^{'}$,
$X_1$ and $X_2$ represent the relevant restricted partition functions
for the real space RG of a system containing a $1$-defect.
At the $0$-th level, $A^{'}_{0} = h$ (where $h$ is an external field
which gives rise to this defect) and $C^{'}_{0} = {X}_{1,0} ={X}_{2,0}
= 0$. The recursion equations for $A^{'}$, $C^{'}$, $X_1$ and $X_2$
(scaled by $Z$ as usual) are as follows:
\bea
{\tilde A}^{'}_{r+1} &=& \frac{{\tilde A}^{'}_{r} (1 + 2 {\tilde F}_r + 2
  {\tilde F}^2_r) + (2+n) {\tilde F}_r^2 {\tilde C}_r^{'} + 2  {\tilde G}_r
{\tilde F}_r {\tilde X}_{1r} 
+ {\tilde F}_r^2 {\tilde X}_{2r}
}{1 + n {\tilde F}_r^3} \\
{\tilde C}^{'}_{r+1} &=& \frac{{\tilde F}^2_r{\tilde A}^{'}_{r} + 3 {\tilde
    F}^2_r {\tilde C}^{'}_{r} +  {\tilde G}_r^2 {\tilde C}^{'}_{r}}{1 + n {\tilde
  F}_r^3} \\
{\tilde X}_{1,{r+1}} &=&  \frac{ 2 {\tilde G}_r {\tilde F}_r {\tilde A}^{'}_{r} +
4 {\tilde G}_r {\tilde F}_r {\tilde C}^{'}_{r} + (2 {\tilde F}_r + 
3 {\tilde F}_r^2 +  {\tilde G}_r^2) {\tilde X}_{1,{r}} 
+ 2 {\tilde G}_r {\tilde F}_r {\tilde X}_{2,{r}}
}{1 + n {\tilde F}_r^3}
\\
{\tilde X}_{2,{r+1}} &=& \frac{6 {\tilde G}_r^2{\tilde C}^{'}_{r} +   
6 {\tilde G}_r {\tilde F}_r {\tilde X}_{1,{r}} + 
(3 {\tilde F}_r^2 + 3 {\tilde G}_r^2){\tilde X}_{2,{r}}}{1 + n {\tilde F}_r^3}
\eea

\begin{table}
\begin{center}
\large{
\begin{tabular}{|c|c|c|c|c|c|c|} \hline 
$n$ & $-2$ & $-1$ & $0$ & $1$ & $2$ &  $\infty$  \\ \hline
$\lambda_1^{\rm bicon}$ & $10/3$ & $3.23$ & $3.21$ & $3.16$ & $3.13$ & $3$ \\ \hline
\end{tabular}
}
\end{center}
\caption{Some values of $\lambda_1^{\rm bicon}$ versus $n$ associated with the $1$-defect.}
\end{table}

Solving the $4 \times 4$ matrix obtained by linearizing the above
equations about the fixed points in Eq.~(\ref{newline}), the largest
eigenvalue $\lambda_1^{\rm bicon}$, gives the scaling dimension
$y_1^{\rm bicon}$ associated with a $1$-defect: 
\bea y_1^{\rm bicon} = {\rm ln}~{\lambda_1^{\rm bicon}}/{{\rm ln}~2}.  
\eea 
Some values of
$\lambda_1^{\rm bicon}$ versus $n$ are given in Table $1$. Note that
upon approaching the point $(n = -2, \omega = 2, \omega_3 = 0)$, we
have $y_1^{\rm bicon} \rt y_1$, i.e., we recover the scaling exponent
of the $1$-string defect in the loop model, as expected.

{\it $2$-defect dimension}: Again the relevant restricted partition 
functions are shown in Fig.~\ref{net4a} by representative symbols 
$X_3$ and $X_4$. The recursions for the latter scaled by $Z$ are as 
follows: 
\bea
{\tilde X}_{3,r+1} &=& \frac{{\tilde X}_{3,r} (2 {\tilde F}_r + 3
  {\tilde F}^2_r)}{1 + n {\tilde F}_r^3} \\
{\tilde X}_{4,r+1} &=& \frac{2 {\tilde X}_{3,r}{\tilde F}_{r} {\tilde G}_r + 
(3 {\tilde F}^2_r  + {\tilde G}_r^2) {\tilde X}_{4,r}}{1 + n {\tilde
  F}_r^3} 
\eea
Assuming $X_{3,r} \sim (\lambda_2^{\rm bicon})^{r}$ the scale factor 
$\lambda_2^{\rm bicon} = (2 \omega_*^{-1} + 3 \omega_*^{-2})/(1 + 
n \omega_*^{-3})$ and independent of $\omega_{3}$. Putting the critical 
point value $\omega_* = 2$ (Eq.~(\ref{newline})), we get:
\bea
\lambda_2^{\rm bicon} = \frac{14}{8+n}.
\eea

{\it $3$-defect dimension}: For this defect the relevant restricted
partition function is shown in Fig.~\ref{net4a} and is represented by
the symbol $X_5$. The recursion relation for ${\tilde X}_{5} = X_5/Z$
is: \bea {\tilde X}_{5,r+1} = \frac{3 {\tilde F}_r^2 {\tilde X}_{5,r}
}{1 + n {\tilde F}_r^3} \eea and the corresponding scale factor on the
critical line (Eq.~(\ref{newline})) is \bea \lambda_3^{\rm bicon} =
\frac{6}{n + 8}.  \eea Thus for $n > -2$, the scale factor
$\lambda_3^{\rm bicon} < 1$ and hence the $3$-defect is irrelevant on
the entire new critical line (Eq.~(\ref{newline})). This was certainly
to be expected, since it is exactly this $3$-defect that induces the
flow from the $\omega_3=0$ line to the new critical line.
At the point $(n = -2, \omega = 2, \omega_3 =
0)$, we have $\lambda_3^{\rm bicon} = 1$ as expected, since 
$\lambda_3 = 1$ for the loop model at that point.

\section{The loop model with local bending energy}
\label{bend}

The problem of self-avoiding polymers with bending energy has been of
long standing interest, and was first introduced by Flory \cite{Flory}.  With
high energy cost for bending, the polymer is in an ordered state (with
minimal bending), while reducing the energy cost leads to a disordered
(but critical) state. The nature of the phase transition separating the
two phases was unclear for a long time, and finally it was shown
recently that for a compact (i.e., space filling) polymer on a two-dimensional
square lattice the transition is continuous
\cite{kondev04a,kondev04b}. The latter works actually dealt with the
full loop model and obtained the relevant results for the polymer by
taking the $n \rt 0$ limit. In a similar spirit we would look at the
loop model for general $n$ with local bending energy on $3$-simplex
fractal lattice; unlike \cite{kondev04a,kondev04b} our loops are not 
compact.  We note that the $n \rt 0$ limit has already been
studied earlier in \cite{maritan} where it was found that the bending energy 
is irrelevant and no new fixed points appear in the extended phase space.
We show below that the same is true for general $n$.  

\begin{figure}
\includegraphics[width=10cm]{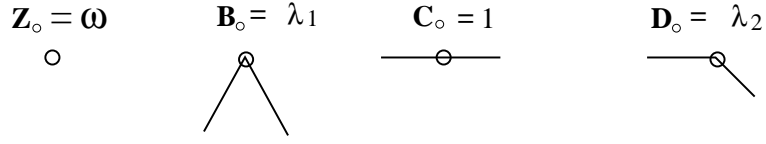}
\caption{The different local configurations for a vertex is shown. For 
a covered vertex the loop can either go straight or bend in two ways.}
\label{bend1}
\end{figure}

\begin{figure}
\includegraphics[width=10cm]{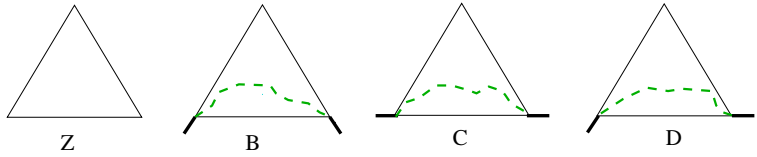}
\caption{Possible restricted partition functions are represented schematically
above. Note that the external legs covered by loops are distinct in all of 
them.}
\label{zabc}
\end{figure}

The model to be studied here is defined as 
\beq 
Z_{\rm loop} = \sum_{\cal C} n^{\cal L} \omega^{\cal V} \lambda_1^{{\cal V}_1}
\lambda_2^{{\cal V}_2}. 
\label{Zbend1}
\eeq On a $3$-simplex fractal lattice, at any vertex, a loop can go
straight, or bend by $\frac{2\pi}{3}$, or $\frac{\pi}{3}$ (see
Fig.~\ref{bend1}).  Accordingly we define local vertex weights $1$,
$\lambda_1$, $\lambda_2$ respectively, for the three
cases. Correspondingly, the number of vertices covered with loops
having $\frac{2\pi}{3}$ and $\frac{\pi}{3}$ bends are respectively
${{\cal V}_1}$ and ${{\cal V}_2}$. The remainder of the symbols in
Eq.~(\ref{Zbend1}) are as in Eq.~(\ref{Zloop}) for the partition function of
the loop model. For doing the real space RG in this case, we require
four restricted partition functions with distinct external legs
covered by loops as shown in Fig.~\ref{zabc}.  These are further
represented by symbols $Z,B,C$ and $D$, and at level $0$, the values
of these partition functions are shown in Fig.~\ref{bend1}. Note that
for $\lambda_1 = \lambda_2 = 1$, the diagrams $B$, $C$ and $D$ are equal.

\begin{figure}
\includegraphics[width=8cm]{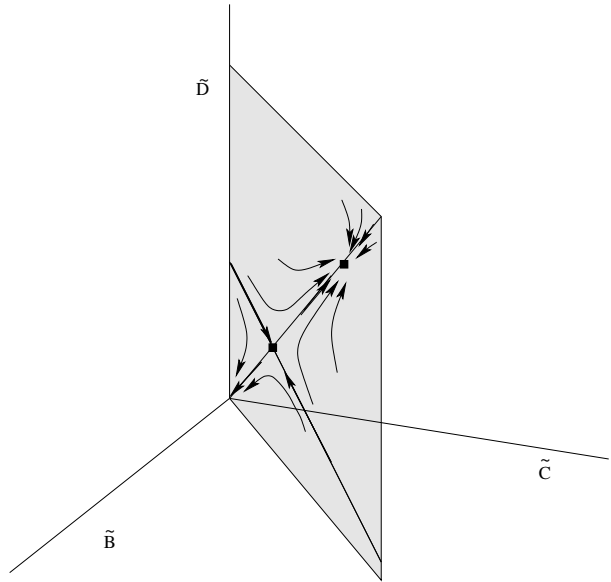}
\caption{The RG flow shown on a plane containing the $(1,1,1)$ direction 
in the ${\tilde B},{\tilde C},{\tilde D}$ space for a fixed $n$.}
\label{flow}
\end{figure}

The recursion relations for the restricted partition functions are 
\bea
Z_{r+1}&=& Z_r^3 + n B_r^3 \nonumber \\
\label{Zbend}
B_{r+1}&=&B_r C_r^2 + Z_r D_r^2 \nonumber \\ 
\label{Bbend} 
C_{r+1}&=&B_r D_r^2 + Z_r C_r^2 \nonumber \\ 
\label{Cbend} 
D_{r+1}&=&B_r C_r D_r + Z_r C_r D_r, 
\label{Dbend} 
\eea
with $Z_0=\omega$, $B_0=\lambda_1$, $C_0=1$ and $D_0=\lambda_2$ 
(see Fig.~\ref{bend1}). If we scale $B$, $C$ and $D$ by $Z$, we get 
\bea
{\tilde B}_{r+1}&=&\frac{B_{r+1}}{Z_{r+1}}=\frac{{\tilde B}_r {{\tilde C}_r}^2 + {{\tilde D}_r}^2}{1 + n {{\tilde B}_r}^3}  
\label{B1} \\
{\tilde C}_{r+1}&=&\frac{C_{r+1}}{Z_{r+1}}=\frac{{\tilde B}_r {{\tilde D}_r}^2 + {{\tilde C}_r}^2}{1 + n {{\tilde B}_r}^3}  
\label{C1} \\
{\tilde D}_{r+1}&=&\frac{D_{r+1}}{Z_{r+1}}=\frac{{\tilde B}_r {\tilde C}_r {\tilde D}_r + {\tilde D}_r {\tilde C}_r}{1 + n {{\tilde B}_r}^3}  
\label{D1} 
\eea From the above Eqs.~(\ref{B1})--(\ref{D1}) we get the following
fixed points (assuming $\lambda_1~{\rm and}~\lambda_2 \neq0$): 
\bea 
 {\tilde B}_* = {\tilde C}_* = {\tilde D}_* ~{\rm and}~ 1+n {\tilde C}_{*}^{3}= {\tilde C}_{*}^{2} + {\tilde C}_{*}.
 \label{Frelation} 
\eea 
Comparing this with
Eq.~(\ref{fixed}) we see that non-zero $\lambda_1$ and $\lambda_2$
have no new effect on the loop model and no new critical points come into
being. The irrelevance of bending energy is shown in gray in Fig.~\ref{flow}. 
The figure shows two fixed points in the $(1,1,1)$ direction
in the ${\tilde B},{\tilde C},{\tilde D}$ space for fixed $n$. 
The attractive fixed point and the repulsive fixed point in the 
$(1,1,1)$ direction are exactly the same as two points for a 
fixed $n$ (belonging to the two branches) in Fig.~\ref{branch}. 
The RG flow diagram on a plane 
containing the $(1,1,1)$ line is shown in Fig.~\ref{flow}.

The irrelevance of bending rigidity is not found for all fractal
lattices. In \cite{maritan}, as well as in our calculation above,
exponents do not get affected on the $3$-simplex lattice. Similarly,
in \cite{klein} it was shown that there is no effect of the strength
of self-interaction on the swelling exponent of a self-avoiding walk
on the Sierpinski gasket.  However on another fractal, namely the
branching Koch curve (BKC) the exponents do change. This was shown for
the self-avoiding polymers \cite{maritan} and recently the full loop model 
for general $n$ \cite{supravat} on BKC.

\section{Conclusion}
\label{conclude}

In summary, we have studied the $n$-vector model on a fractal lattice
with dimension $d<2$.  We have shown that---just like its counterpart
in $d=2$ \cite{Nienhuis82}---it has two (physical) critical branches,
referred to as `dilute' and `dense' loops. However, while these
branches exist for $0 \leq n \leq 2$ in $d=2$, on the 3-simplex
lattice they are constrained to $0 \leq n \leq 1$. We have explicitly
characterized the critical behavior of this model, in terms of three
$k$-string and one thermal exponent.

We note that the dense phase exists on this fractal for $n<1$, and
the dense phase exponents have a nontrivial dependence on $n$. This is
in contrast to the expected behavior of the $n$-vector model for $d>
2$, $n>1$. For $n>1$ and $d>2$, the low temperature phase of the
$n$-vector model in zero field is characterized by a spontaneous
magnetization. The entire low-temperaure phase is critical, with
infinite correlation length, because of the presence of Goldstone
modes in the system.  There are $n-1$ Goldstone modes, as there are
$(n-1)$ directions orthogonal to the magnetization. For each of these
modes, the excitation energy for an excitation of wave number $k$
varies as $k^2$. Using this, and the fact that the spin wave
approximation becomes asymptotically exact at low temperatures, one
can easily deduce results like the spontatenous magnetization at low
temperature $T$ varies as $1 - A (n-1) T^{d/2}$ for general $n$.
Also, the mean energy at temperature $T$ varies as $E(T) = E(T=0) +
B (n-1) T^{d/2 +1}$.  In all these cases, only the amplitudes are 
proportional to $(n-1)$, and the critical exponents do not
depend on $n$. However, for the fractal we studied, the dense phase
exists only for $n<1$. In this case, if we do a naive analytic
continuation in $n$, the number of Goldstone modes becomes
negative. Then, for $n <1$, magnetization density becomes $>1$, and
$E(T)$ decreases with increasing $T$. Clearly the argument that the
exponents of the phase are determined by Goldstone modes no longer
applies.

We have further generalized the loop model into the `biconnected clusters
model' which allows for a finite density of 3-valent vertices. This model
was found to support a new line of $n$-dependent critical points, emanating
from the $n=-2$ point on the dilute branch in the original model. At this
point, the 3-string defect is a marginal perturbation, which is clearly a
necessary requirement for the emergence of a new critical line in the
generalized model.

An interesting line of future research would be to study models
allowing for 3-valent vertices in two dimensions. Also in $d=2$, the
3-string defect is marginal at the $n=-2$ point on the dilute branch.
Within conformal field theory, the latter point is a theory of
symplectic fermions with central charge $c=-2$. One might speculate
that in this case the perturbation could be {\it exactly} marginal,
and hence generate a line of $c=-2$ theories with continuously varying
critical exponents. Such theories were recently shown to exist
\cite{Candu}, and can be produced \cite{Candu} by perturbing the
$c=-2$ point on the {\it dense} branch (i.e., $n=0$) by a finite
density of {\it six}-valent vertices.

\subsection*{Acknowledgements}
 
The authors acknowledge grant no.~3404--2 of the ``Indo-French Center
for the Promotion of Advanced Research (IFCPAR)/Centre Franco-Indien
pour la Promotion de la Recherche Avanc\'ee (CEFIPRA)'' for financial
support. Deepak Dhar would also like to acknowledge support of
Department of Science and Technology, Government of India, through a
J. C. Bose fellowship.


\begin{thebibliography}{99}

\bibitem{Mukamel}    E.~Domany, D.~Mukamel, B.~Nienhuis, and A.~Schwimmer,
                     Nucl.~Phys.~B {\bf 190}, 279 (1981).
\bibitem{Nienhuis82} B.~Nienhuis, Phys.~Rev.~Lett.~{\bf 49}, 1062 (1982).
\bibitem{Baxter86}   R.~J.~Baxter, J.~Phys.~A {\bf 19}, 2821 (1986).
\bibitem{deGennes}   P.~G.~de Gennes, Phys.~Lett.~A {\bf 38}, 339 (1972).
\bibitem{Zinnbook}   J.~Zinn-Justin, {\em Quantum field theory and
                     critical phenomena} (Clarendon Press, Oxford, 1989).
\bibitem{Saleur86}   H.~Saleur, J.~Phys.~A {\bf 19}, L807 (1986).
\bibitem{Batchelor88}M.~T.~Batchelor and H.~W.~J.~Bl\"ote,
                     Phys.~Rev.~Lett.~{\bf 61}, 138 (1988);
                     Phys.~Rev.~B {\bf 39}, 2391 (1989).
\bibitem{DupSal87}   B.~Duplantier and H.~Saleur,
                     Nucl.~Phys.~B {\bf 290}, 291 (1987).
\bibitem{Nienhuis92} S.~O.~Warnaar, M.~T.~Batchelor and B.~Nienhuis,
                     J.~Phys.~A {\bf 25}, 3077 (1992).
\bibitem{Kondev98}   J.~L.~Jacobsen and J.~Kondev,
                     Nucl.~Phys.~B {\bf 515}, 701 (1998).
\bibitem{Batchelor94}M.~T.~Batchelor, J.~Suzuki and C.~M.~Yung,
                     Phys.~Rev.~Lett.~{\bf 73}, 2646 (1994).
\bibitem{Kondev95}   J.~Kondev, J.~de Gier and B.~Nienhuis,
                     J.~Phys.~A {\bf 29}, 6489 (1996).
\bibitem{guo}        W. Guo, H.~W.~J.~Bl\"ote, and F. Y. Wu, 
                     Phys. Rev. Lett. {\bf 85}, 3874 (2000).   
\bibitem{das}        D. Das and J. L. Jacobsen, 
                     J. Phys. A {\bf 37}, 1 (2004).
\bibitem{read}       J.L. Jacobsen, N. Read and H. Saleur, Phys. 
                     Rev. Lett. {\bf 90}, 090601 (2003). 
\bibitem{Fendley}    P. Fendley and E. Fradkin, Phys. Rev. B {\bf 72},
                     024412 (2005); P. Fendley, {\em Topological order
                     from quantum loops and nets}, arXiv:0804.0625.
\bibitem{Zinn}       J.L. Jacobsen and P. Zinn-Justin, J. Phys. A
                     {\bf 35}, 2131 (2002); Phys. Rev. E {\bf 66},
                     055102(R) (2002).
\bibitem{Deng}       Y. Deng, H.W.J. Bl\"ote and B. Nienhuis,
                     Phys. Rev. E {\bf 69}, 026114 (2004).
\bibitem{kenevic}    Y. Gefen {\it et. al.} Phys. Rev. Lett. {\bf 47}, 
                     1771 (1981); M. Kne$\breve{\rm z}$evi$\acute{\rm c}$, 
                     Ph.d. thesis, at L'$\acute{\rm E}$cole Normale 
		     Sup$\acute{\rm e}$riure (1986). 
\bibitem{nelson}     D.R. Nelson and M.E. Fisher, Ann. Phys. (N.Y.) {\bf 91},
                     226 (1975).   
\bibitem{dhar2}      D. Dhar, J. Math. Phys. {\bf 19}, 5 (1978).
\bibitem{klein}      D.J. Klein and W.A. Seitz, J. Physique Lett. {\bf 45},
                     L241 (1984).
\bibitem{Chang}      I.S. Chang and Y. Shapir, J. Phys. A: Math. Gen. {\bf 21},
                     L903 (1988).
\bibitem{kenevic1}   M. Kne$\breve{\rm z}$evi$\acute{\rm c}$  and 
                     B.W. Southern, Phys. Rev. B {\bf 34}, 4966 (1986).  
\bibitem{kenevic2}   M. Kne$\breve{\rm z}$evi$\acute{\rm c}$ and 
                     J. Vannimenus, Phys. Rev. Lett. {\bf 56}, 
                     1591 (1986). 
\bibitem{abhishek}   A. Dhar and D. Dhar, Phys. Rev. E {\bf 55}, R2093 (1997). 
\bibitem{sumedha}    Sumedha and D. Dhar, J. Stat. Phys. {\bf 125}, 55 (2006).
\bibitem{maritan}    A. Giacometti and A. Maritan, J. Phys. A: Math. Gen. 
                     {\bf 25}, 2753 (1992).
\bibitem{dhar3}      D. Dhar and Y. Singh, in {\em Statistics of 
                     linear polymers in disordered media}, Ed. B. K. 
                     Chakrabarti, (Elsevier, 2005).
\bibitem{Cardy}      J.~Cardy, {\em Scaling and Renormalization in  
                     Statistical Physics} (Cambridge Univ. Press, 2002).
\bibitem{supravat}   S.~Dey and D.~Das, unpublished.  
\bibitem{saleur}     H. Saleur and B. Duplantier, Phys. Rev. Lett. {\bf 58},
                     2325 (1987). 
\bibitem{vani_df}    M. Kne$\breve{\rm z}$evi$\acute{\rm c}$ and 
                     and J. Vannimenus, J. Phys. A {\bf 20}, 
                     1215 (1987). 
\bibitem{kondev_loop} J. Kondev, C.L. Henley and D. G. Salinas, 
                     Phys. Rev. E {\bf 61}, 104 (2000).     
\bibitem{Flory}      P.~J.~Flory, Proc.~R.~Soc.~London, Ser.~A{\bf 234}, 
                     60 (1956).
\bibitem{kondev04a}  J.~L.~Jacobsen and J.~Kondev, Phys.~Rev.~Lett.~{\bf 92}, 
                     210601 (2004). 
\bibitem{kondev04b}  J.~L.~Jacobsen and J.~Kondev, Phys.~Rev.~E {\bf 69}, 
                     066108 (2004) and references therein.
\bibitem{Candu}      C.~Candu, J.~L.~Jacobsen and H.~Saleur, in preparation.

\end{thebibliography}
\end{document}